\documentclass[5p,times,twocolumn,preprint]{elsarticle}
\usepackage{graphicx}
\usepackage{lineno}
\usepackage{xcolor}
\usepackage{subfig}
\usepackage[colorlinks=true,
            linkcolor=blue,
            citecolor=blue,
            urlcolor=blue]{hyperref}
\usepackage{amsmath}

\begin{document}

\begin{frontmatter}

\title{Beam Test Characterization of Silicon Microstrip Detector Flight-Model Ladders for the AMS-02 Upgrade}

\author[1,2]{Dexing~Miao}
\author[7]{Giovanni~Ambrosi}
\author[7]{Mattia~Barbanera}
\author[1]{Baasansuren~Batsukh}
\author[1,2]{Hengyi~Cai}
\author[1]{Mengke~Cai}
\author[3]{Xudong~Cai}
\author[1,2]{Yuman~Cai}
\author[6]{Yuan-Hann~Chang}
\author[1]{Shanzhen~Chen}
\author[6]{Hsin-Yi~Chou}
\author[1]{Xingzhu~Cui}
\author[1,2]{Mingyi~Dong}
\author[7]{Matteo~Duranti}
\author[1]{Ke~Gong}
\author[1,2]{Mingjie~Feng}
\author[13]{Valerio~Formato}
\author[1,2]{Yisheng~Fu}
\author[4,5]{Daojin~Hong}
\author[7]{Maria~Ionica}
\author[1,2]{Xiaojie~Jiang}
\author[7]{Yaozu~Jiang}
\author[1]{Liangchenglong~Jin}
\author[1,9]{Shengjie~Jin}
\author[3]{Vladimir~Koutsenko}
\author[1,2]{Qinze~Li}
\author[1,10]{Tiange~Li} 
\author[1,2]{Zuhao~Li}
\author[6]{Chih-Hsun~Lin}
\author[1,2]{Changcheng~Liu}
\author[4]{Cong~Liu}
\author[11]{Hanbing~Liu}
\author[4]{Pingcheng~Liu\corref{cor1}}\ead{pingcheng.liu@iat.cn}
\author[8]{Alberto~Oliva}
\author[1,2]{Ji~Peng}
\author[1]{Wenxi~Peng}
\author[1]{Rui~Qiao}
\author[1,2]{Shuqi~Sheng}
\author[1,2]{Tianyu~Shi}
\author[7]{Gianluigi~Silvestre}
\author[1]{Zetong~Sun}
\author[1]{Congcong~Wang}
\author[1]{Feng~Wang}
\author[4]{Hongbo~Wang}
\author[11]{Zhijie~Wang}
\author[5]{Zibing~Wu}
\author[1,12]{Zhiyu~Xiang}
\author[4]{Suyu~Xiao}
\author[4,5]{Weiwei~Xu}
\author[1,2]{Zixuan~Yan}
\author[1,2]{Haotian~Yang}
\author[1]{Sheng~Yang}
\author[1,2]{Yuhang~You}
\author[1]{Xuhao~Yuan}
\author[1,2]{Yuan~Yuan}
\author[1]{Fengze~Zhang}
\author[1]{Xiyuan~Zhang}
\author[1]{Zijun~Xu\corref{cor1}}\ead{xuzj@ihep.ac.cn}
\author[1]{Jianchun~Wang}

\cortext[cor1]{Corresponding author}

\address[1]{Institute of High Energy Physics, Beijing 100049, China}
\address[2]{University of Chinese Academy of Sciences, Beijing 100049, China}
\address[3]{Massachusetts Institute of Technology (MIT), Cambridge, Massachusetts 02139, USA}
\address[4]{Shandong Institute of Advanced Technology, Jinan
250100, China}
\address[5]{Shandong University (SDU), Jinan, Shandong 250100, China}
\address[6]{Institute of Physics, Academia Sinica, Nankang, Taipei, 11529, Taiwan}
\address[7]{INFN Sezione di Perugia, 06100 Perugia, Italy}
\address[8]{INFN Sezione di Bologna, 40126 Bologna, Italy}
\address[9]{Northwest Normal University, Lanzhou 730070, China}
\address[10]{Hunan University, Changsha 410082, China}
\address[11]{Lanzhou University, Lanzhou 730000, China}
\address[12]{Central South University, Changsha 410083, China}
\address[13]{INFN Sezione di Roma2, 00133 Roma, Italy}
\begin{abstract}
The AMS-02 experiment plans to install a new silicon microstrip tracker layer (Layer-0) on top of the existing detector, increasing the cosmic-ray acceptance by a factor of 3. Layer-0 employs a design in which multiple silicon microstrip detectors (SSDs) are connected in series to form long detector ladders. We present a detailed performance study of the flight-model ladders using a 350~GeV mixed hadron beam at the CERN SPS. The study focuses on the following aspects:
(i) the performance of ladders with different numbers of SSDs, for which the intrinsic spatial resolution at normal incidence varies from $9.5~\mu\mathrm{m}$ to $11.4~\mu\mathrm{m}$ for ladders composed of 8 to 12 SSDs;
(ii) the response consistency for particles impacting on the \emph{Head} and \emph{Tail} regions of the ladder; and
(iii) the dependence of the detector performance on the particle incidence angle.
\\
\noindent Keywords: AMS-02 Layer-0, silicon microstrip detector, test beam, spatial resolution

\end{abstract}
\end{frontmatter}

\section{Introduction }
\label{sec:intro}

The Alpha Magnetic Spectrometer (AMS-02)~\cite{Lubelsmeyer:2011zz}, operating aboard the International Space Station (ISS), is designed to study the origin and evolution of the Universe by searching for antimatter and dark matter, while performing high-precision measurements of the composition and flux of cosmic rays~\cite{AMS:2021nhj}. Equipped with a permanent magnet and multiple sub-detectors, including a silicon tracker, the AMS-02 detector provides precise measurements of particle charge, momentum, and velocity. The tracker system consists of nine layers of silicon detectors~\cite{Duranti:2013qfz}, which are responsible for measuring hit positions and ionization energy losses for incident particles.

\begin{figure*}[ht]
\centering
\subfloat[\label{fig:L0_on_ams}]{\includegraphics[width=0.45\hsize]{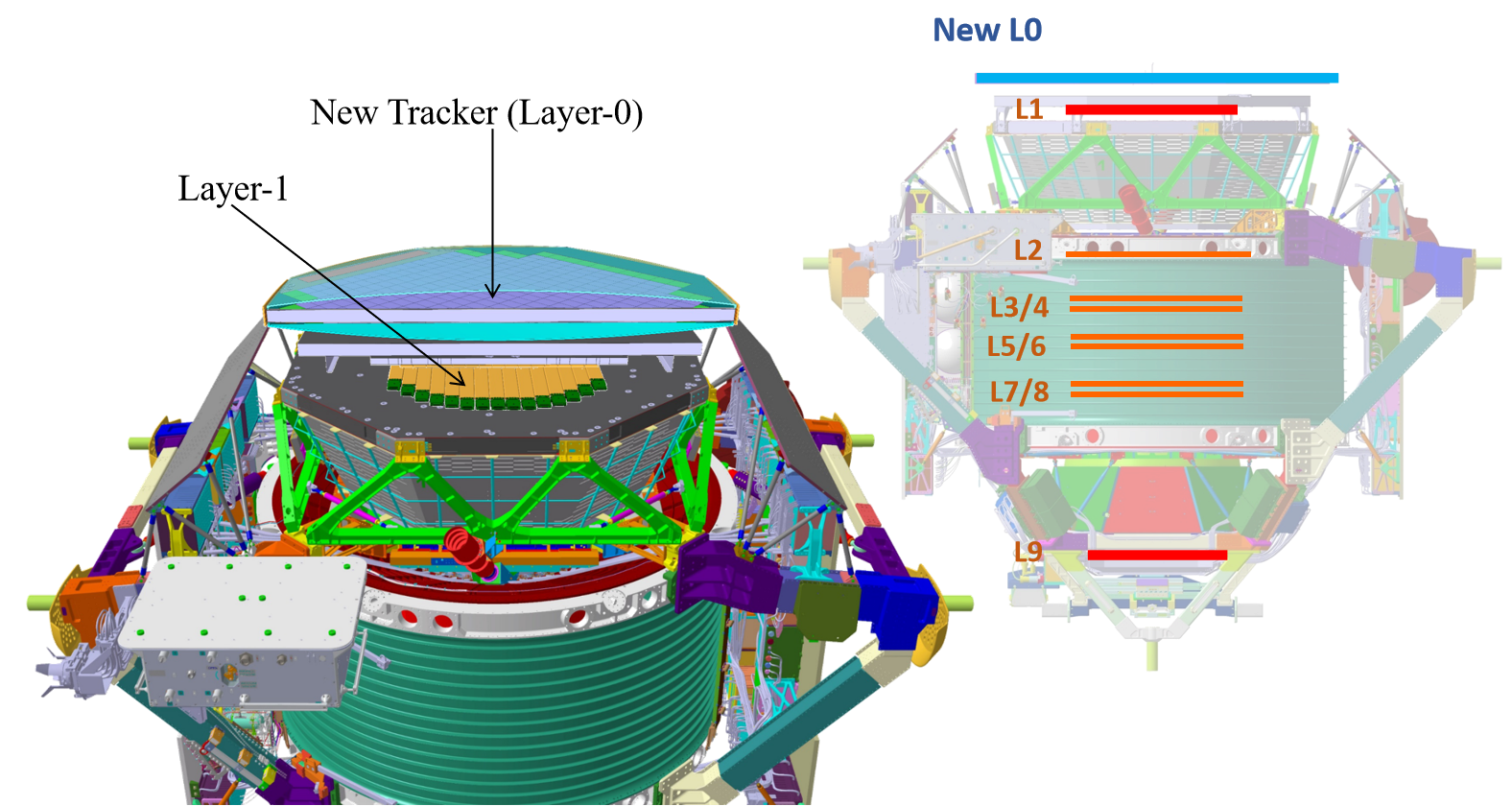}}
\hspace{0.1cm}
\subfloat[\label{fig:L0_plane}]
{\includegraphics[width=0.45\hsize]{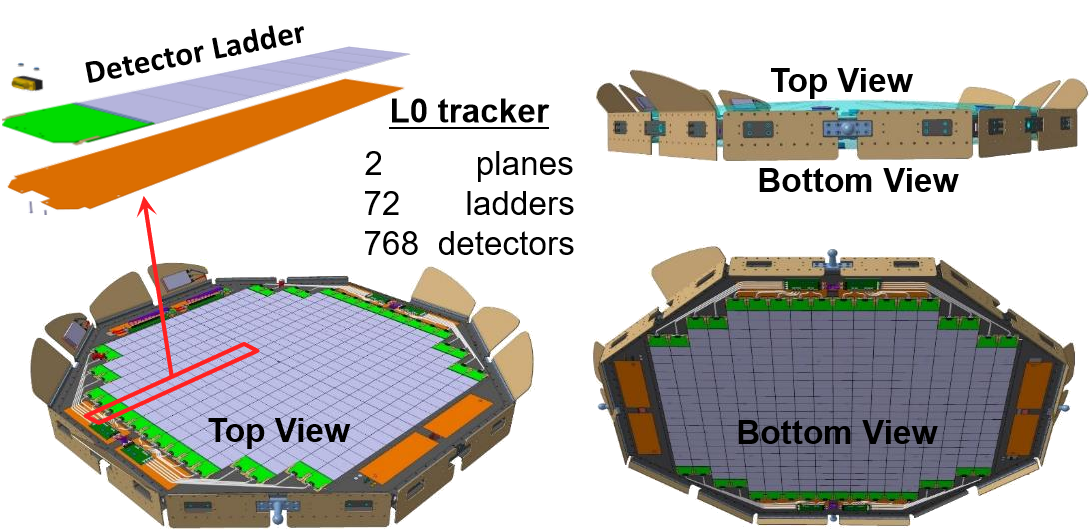}} 
\caption{(a) Left: schematic view of the Layer-0 installed on the top of the AMS-02 detector. Right: side perspective view illustrating the relative size of Layer-0 with respect to the existing silicon tracker layers (L1--L9). (b) Schematic overview of the Layer-0 layout. The upper left panel shows a single detector ladder. The lower left panel presents the top view of the complete Layer-0 assembly, while the right panels display the top and bottom views. The Layer-0 consists of two planes, comprising a total of 72 ladders and 768 silicon microstrip sensors. The images are taken from Ref.~\cite{Gargiulo2025AMS02}.
}
\label{fig:L0_alls}
\end{figure*}

The AMS-02 experiment is planned to operate throughout the full lifetime of the ISS. To enhance the detection acceptance by a factor of 3 and improve the nuclear charge identification capability, a new large-area silicon tracker layer (Layer-0) will be installed on top of AMS-02, as shown in Fig.~\ref{fig:L0_on_ams}. The Layer-0 detector consists of two planes, which are rotated by 45 degrees with respect to each other, and each plane comprises 36 ladders, as illustrated in Fig.~\ref{fig:L0_plane}.  A ladder is the basic detection module of the Layer-0 tracker and is formed by serially connecting 8, 10, or 12 silicon microstrip detectors (SSDs)~\cite{Seidel:2019hty} in a daisy-chain configuration, as shown in Fig.~\ref{fig:ladder}. In total, 76 flight-model (FM) ladders, 72 for installation and 4 spare units, were constructed. Prior to the full plane assembly, a test beam campaign was carried out to characterize the performance of the FM ladders. The study focused on the following aspects:
\begin{itemize}
  \item the performance of ladders with different numbers of SSDs;
  \item the response consistency of the detector along the ladder;
  \item the dependence of the detector performance on the particle incidence angle.
\end{itemize}

This paper is organized as follows. Sec.~\ref{sec2} describes the basic design and configuration of the Layer-0 ladders. Sec.~\ref{sec3} presents the test beam campaign. The performance analysis and results are discussed in Sec.~\ref{sec4}. Finally, a summary is given in Sec.~\ref{sec5}.

\section{Design and configuration of AMS-02 Layer-0 ladders}\label{sec2}
The Layer-0 ladder employs AC-coupled SSDs that are $\mathrm{p}^+\text{-in-}n$ sensors manufactured by Hamamatsu Photonics K.K., with a total active area of $113 \times 80~\mathrm{mm}^2$ and a thickness of $320~\mu\mathrm{m}$. Each SSD consists of 1024 readout strips with a readout pitch of $109~\mu\mathrm{m}$. Three floating strips are implemented between adjacent readout strips to improve the spatial resolution by enhancing charge sharing~\cite{Turchetta:1993vu}.

\begin{figure}[ht]
  \centering
  \includegraphics[width=1\linewidth]{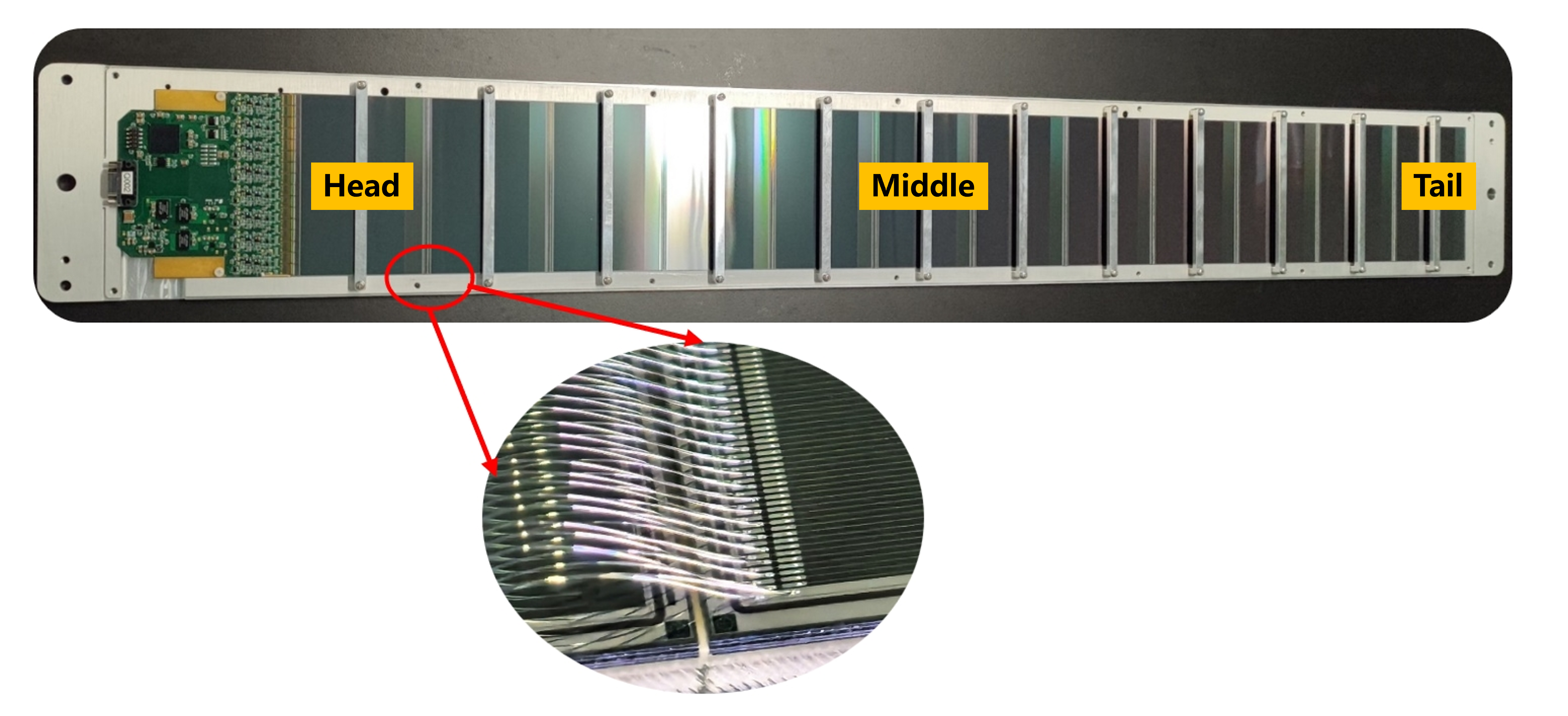}
  \caption{A 12-SSD detector ladder of the AMS Layer-0 tracker. The front-end readout board is located on the left, while the 12 SSDs are connected in series via bonding wires on the right, as shown in the magnified view. The SSD closest to the front-end readout board is defined as the \emph{Head}, the farthest one as the \emph{Tail}, and the central SSD (the sixth one) as the \emph{Middle}; these positions are indicated in the figure.
  \label{fig:ladder} }
\end{figure}

A long silicon detector ladder design is shown in Fig.~\ref{fig:ladder}. The aluminum readout strips of 8/10/12 SSDs are wire-bonded in series to form strips with an effective length of up to 1~m, which are read out by a single electronic channel. Since the majority of the power consumption in a silicon strip detector system comes from the readout electronics, this design reduces the number of readout channels by a factor of $\mathcal{O}(10)$ for the same sensitive area, thus significantly lowering the overall power consumption, which is critical for operating on the ISS.

As illustrated in Fig.~\ref{fig:ladder}, the \emph{Head} SSD of each ladder is wire-bonded to 16 front-end readout chips (IDE1140) on the left side, resulting in a total of 1024 readout channels. This chip provides a gain of $2.6~\mu\mathrm{A/fC}$ in the default configuration, and an equivalent noise charge of $139 e^-$ without load~\cite{IDE1140}. The output signals from the IDE1140 chips are amplified in two stages by an amplifier (AD8031) with a total gain of $6.97\, \mathrm{mV}/\mu\mathrm{A}$, and are subsequently routed to a 14-bit analog-to-digital converter (ADC), LTC2313ITS8-14, with a full-scale input range of $4096~\mathrm{mV}$. The digitized data are then encoded by a field-programmable gate array (FPGA) and transmitted to the back-end. The intrinsic noise level of the readout electronics, measured with no SSD connected, is $\sim \!2$ least significant bits (LSB).
Ladders with a larger number of SSDs exhibit higher noise levels~\cite{Li:2026jxg}, which degrade the signal-to-noise ratio (S/N) and consequently affects the spatial resolution, as discussed in detail in Sec.~\ref{sec_4.1}.

\section{Test beam experiment}\label{sec3}

\begin{figure}[ht]
\centering
\subfloat[\label{fig:test_beam_setup}]{\includegraphics[width=0.66\hsize]{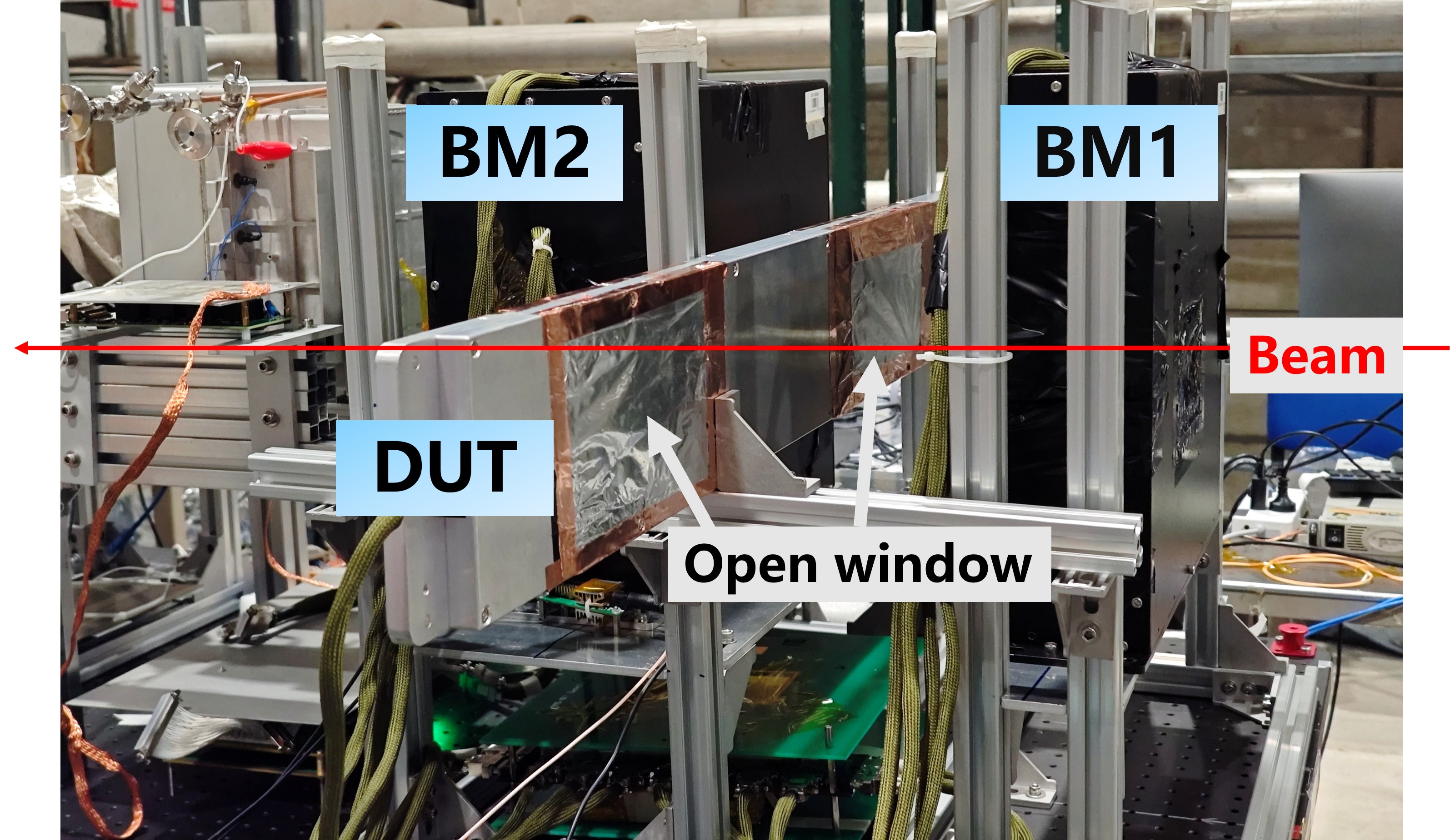}}
\hspace{0.2cm}
\subfloat[\label{fig:test_beam_BM}]
{\includegraphics[width=0.29\hsize]{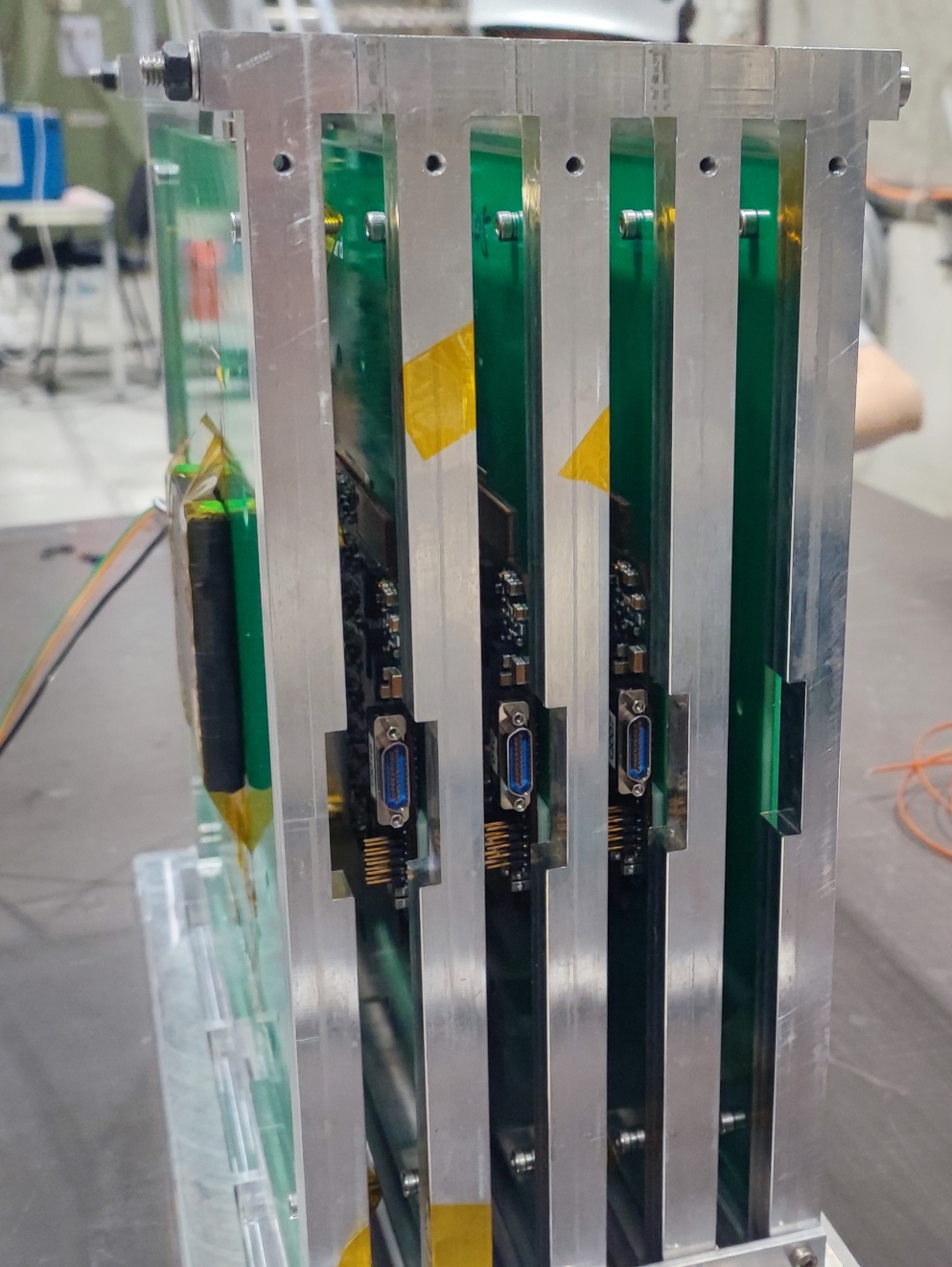}} 
\caption{(a) Photograph of the beam test setup. Along the beam direction, the upstream beam monitor (BM1), the ladder under test, and the downstream beam monitor (BM2) are arranged sequentially. Open windows on the aluminum support box of the ladder to allow the beam to pass through. (b) Photograph showing the internal structure of a beam monitor.}
\label{fig:test_beam_all}
\end{figure}

The beam test presented in this work was carried out in July 2025 at the H4 beamline of the CERN SPS, using a mixed proton and pion beam with a momentum of 350~GeV. A dedicated beam telescope system was constructed, consisting of 12 layers of silicon microstrip detectors. Each telescope layer employs the same SSD and front-end readout electronics as the Layer-0 ladders, but contains a single SSD only. The spatial resolution of a single telescope layer was measured to be $7.2~\mu\mathrm{m}$ for normally incident minimum ionizing particles (MIPs), with a detection efficiency of 99.8\%, as reported in Ref.~\cite{Miao:2025ldv}.

The telescope was divided into two beam monitors (BM), which were positioned upstream and downstream of the ladder under test, referred to as the detector under test (DUT), as illustrated in Fig.~\ref{fig:test_beam_setup}. As shown, windows were opened on the aluminum support box of the ladder and sealed with thin aluminum foil, providing electromagnetic interference shielding and light shielding while minimizing the material budget for beam passing. Each BM comprises six detector layers arranged in an interleaved XY configuration, providing precise tracking and hit position prediction at the DUT. In addition, the coincidence signal of two plastic scintillators coupled to a silicon photomultiplier (SiPM) is used for the trigger~\cite{Zhang:2023mti}. The internal structure of one BM is shown in Fig.~\ref{fig:test_beam_BM}. During data taking, the setup of the DUT was changed according to the specific measurement objectives. 

To compare the performance of ladders with different numbers of SSDs, data were first taken with the beam incident on the \emph{Middle} SSD (the central SSD, cf. Fig.~\ref{fig:ladder}) of ladders composed of 8, 10, and 12 SSDs, respectively. Subsequently, for the 12-SSD ladder, measurements were performed with the beam traversing the \emph{Head} and \emph{Tail} SSDs in order to study the performance consistency along the strip direction over the full ladder length. Finally, data were collected on the \emph{Middle} SSD of the 12-SSD ladder at incidence angles of $10^\circ$, $20^\circ$, $25^\circ$, and $30^\circ$ to investigate the detector performance as a function of the particle incidence angle.

The data-taking rate was limited to about $ 100\,\mathrm{Hz}$ by the simultaneous operation of other parasitic experiments. For each setup, between $100 \,\mathrm{K}$ and $300 \,\mathrm{K}$ events were recorded. Calibrations were performed typically every 1 to 2 hours, to record the pedestal and noise levels of all the readout channels of the BMs and DUT, which could slowly drift with detector temperature.

\section{Performance of the Layer-0 ladders}\label{sec4}
\subsection{Comparison of ladders with different numbers of SSDs}\label{sec_4.1}

\begin{figure}[ht]
\centering
\subfloat[\label{fig:8_10_12_noise}]{\includegraphics[width=0.48\hsize]{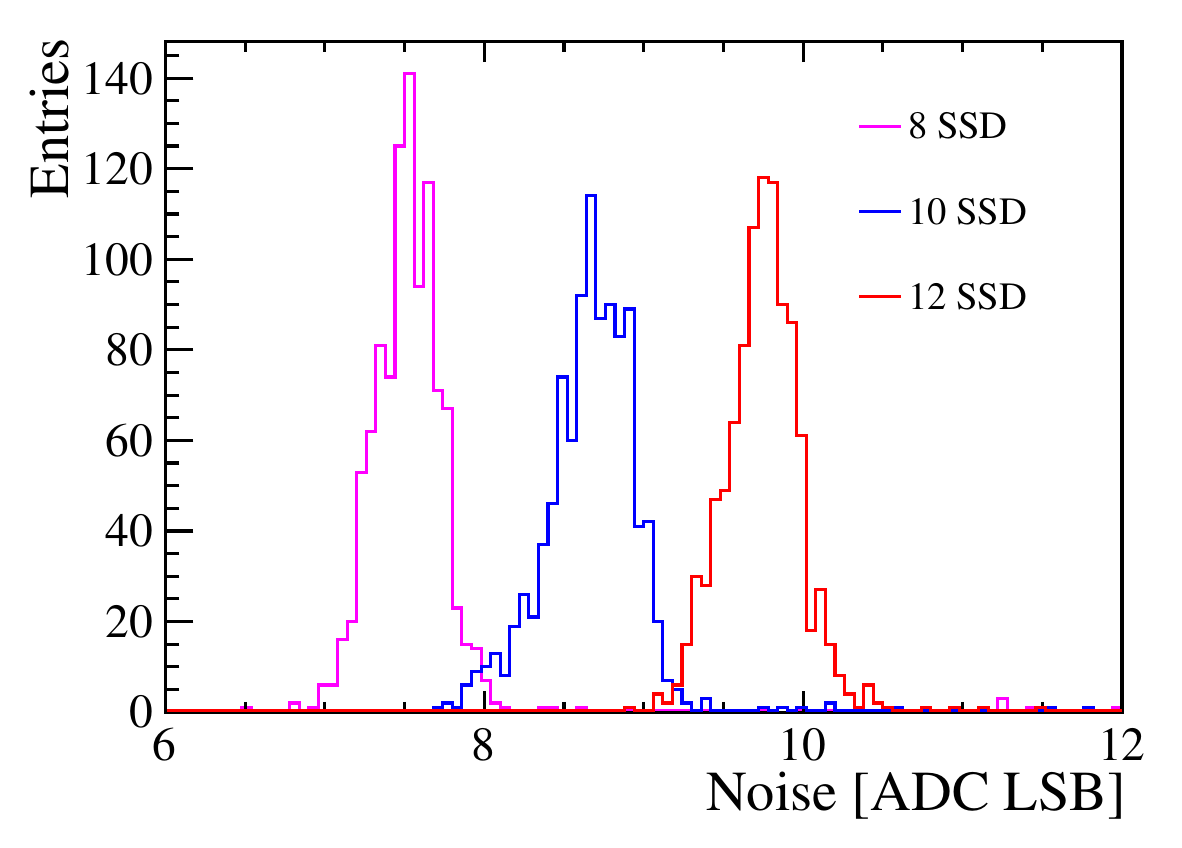}}
\hspace{0.1cm}
\subfloat[\label{fig:8_10_12_MPV}]
{\includegraphics[width=0.48\hsize]{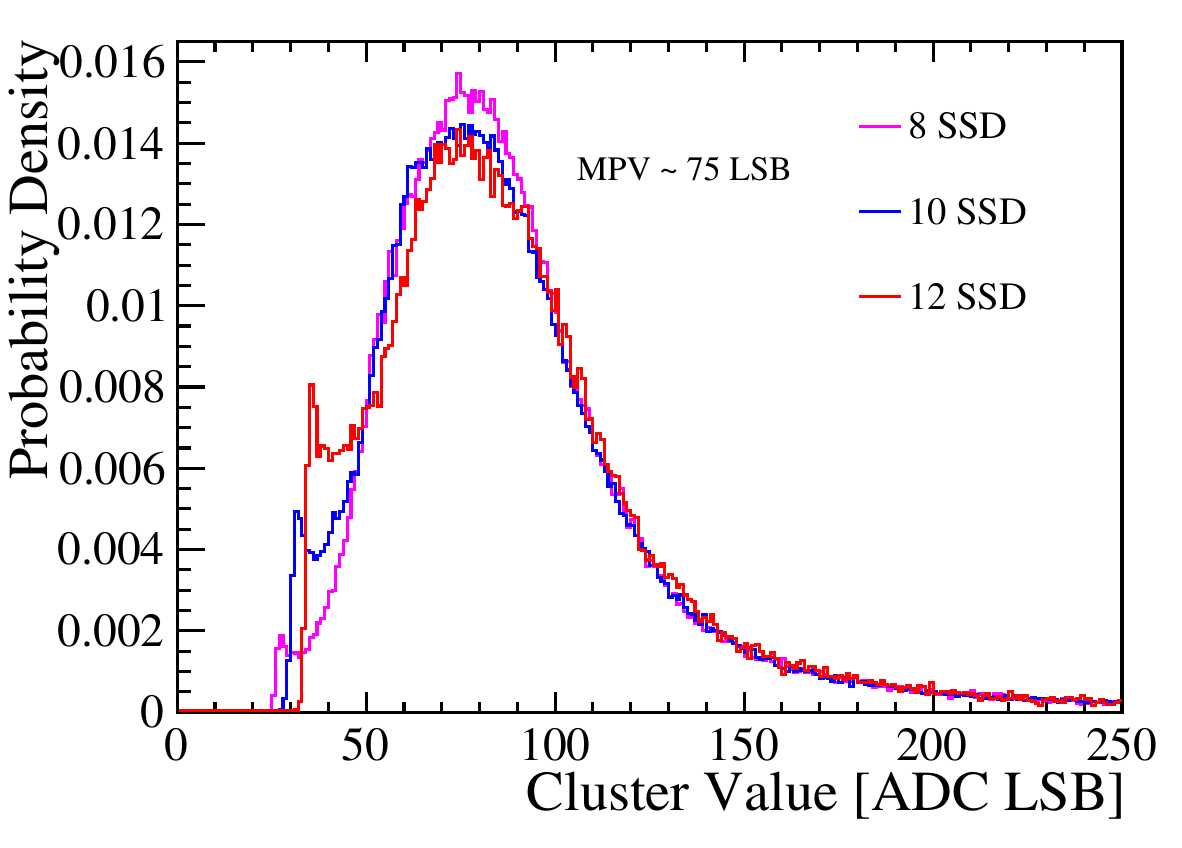}} 
\caption{(a) Intrinsic noise distributions of ladders with 8, 10, and 12 SSDs. The average noise values are 7.6, 8.7, and 9.7~LSB, respectively. (b) Cluster value distributions for the 8-, 10-, and 12-SSD ladders measured when the beam hits the Middle SSD, with the corresponding MPVs all around 75~LSB.}
\end{figure}

We first investigate the noise performance of ladders with different numbers of SSDs. In Fig.~\ref{fig:8_10_12_noise}, the distribution of the intrinsic noise (common-mode noise subtracted~\cite{Abba:2015qka}) values of the 1024 readout channels in each ladder is presented. The noise level increases with the number of SSDs connected in the ladder, with corresponding mean noise values of 7.6, 8.7, and 9.7~LSB for the 8-, 10-, and 12-SSD ladders, respectively. A few channels with abnormally high or low values are marked as bad channels and are not used in the following analysis.

When a charged particle traverses the silicon detector, electron--hole pairs are generated at the p-n junction, collected by the readout strips, and ultimately converted into signals in the corresponding readout channels. As a result, the signal amplitudes of one or several readout channels in the vicinity of the hit position exceed the pedestal level. The group of such channels is referred to as a cluster. For the Layer-0 ladders, clusters are reconstructed using a dual-threshold noise-based method. A readout channel is identified as a seed of a cluster if its ADC value exceeds the pedestal by more than 3.5 times the noise level. Starting from the seed channel, neighboring channels are recursively added to the cluster as long as their ADC values exceed the pedestal by more than 2 times the noise level. The signal value of a readout channel is defined as the ADC value after pedestal and common-mode subtraction~\cite{Miao:2025ldv}. The cluster value is defined as the sum of the signals of all channels belonging to the cluster, while the cluster size is defined as the number of channels contained in the cluster. 

Figure~\ref{fig:8_10_12_MPV} shows the cluster value distributions measured when the beam hits the \emph{Middle} SSD of the 8-, 10-, and 12-SSD ladders. The most probable values (MPVs) are all found to be around 75~LSB, indicating no significant difference in signal amplitude. This indicates that the charge collection efficiency among the ladders with different numbers of SSDs is consistent. The discrepancy in the region between 25 and 50 ADC is due to the different noise levels of the ladders.


\begin{figure}[ht]
\centering
\subfloat[\label{fig:8SSD_residual}]{\includegraphics[width=0.48\hsize]{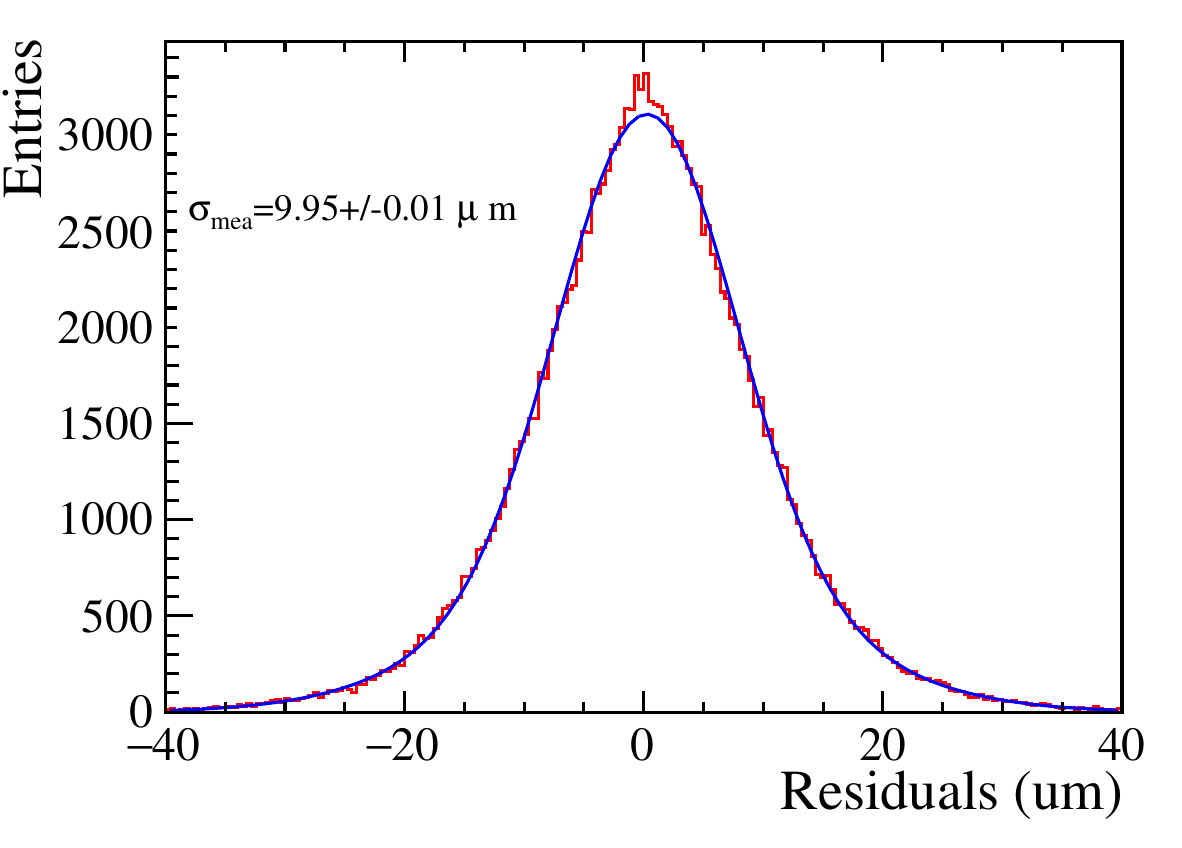}}
\hspace{0.1cm}
\subfloat[\label{fig:10SSD_residual}]
{\includegraphics[width=0.48\hsize]{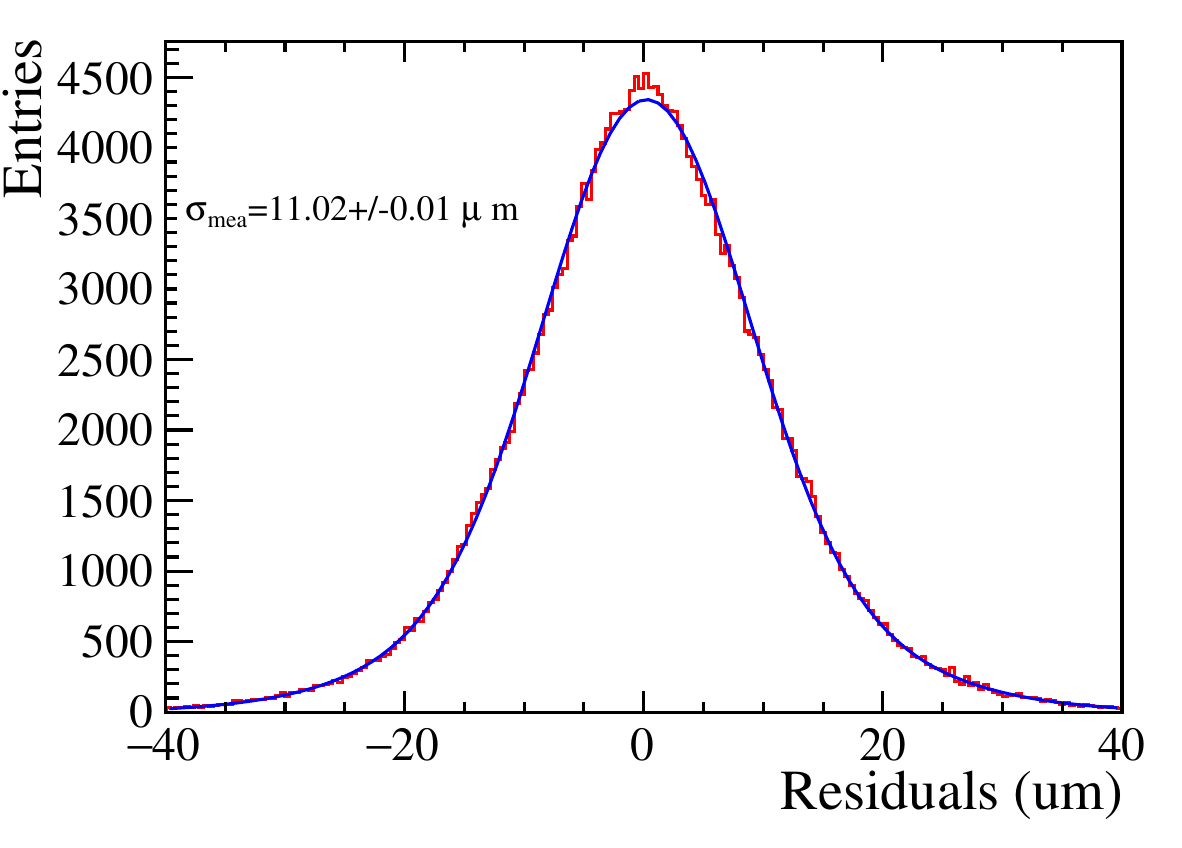}} 
\\
\subfloat[\label{fig:12SSD_residual}]{\includegraphics[width=0.48\hsize]{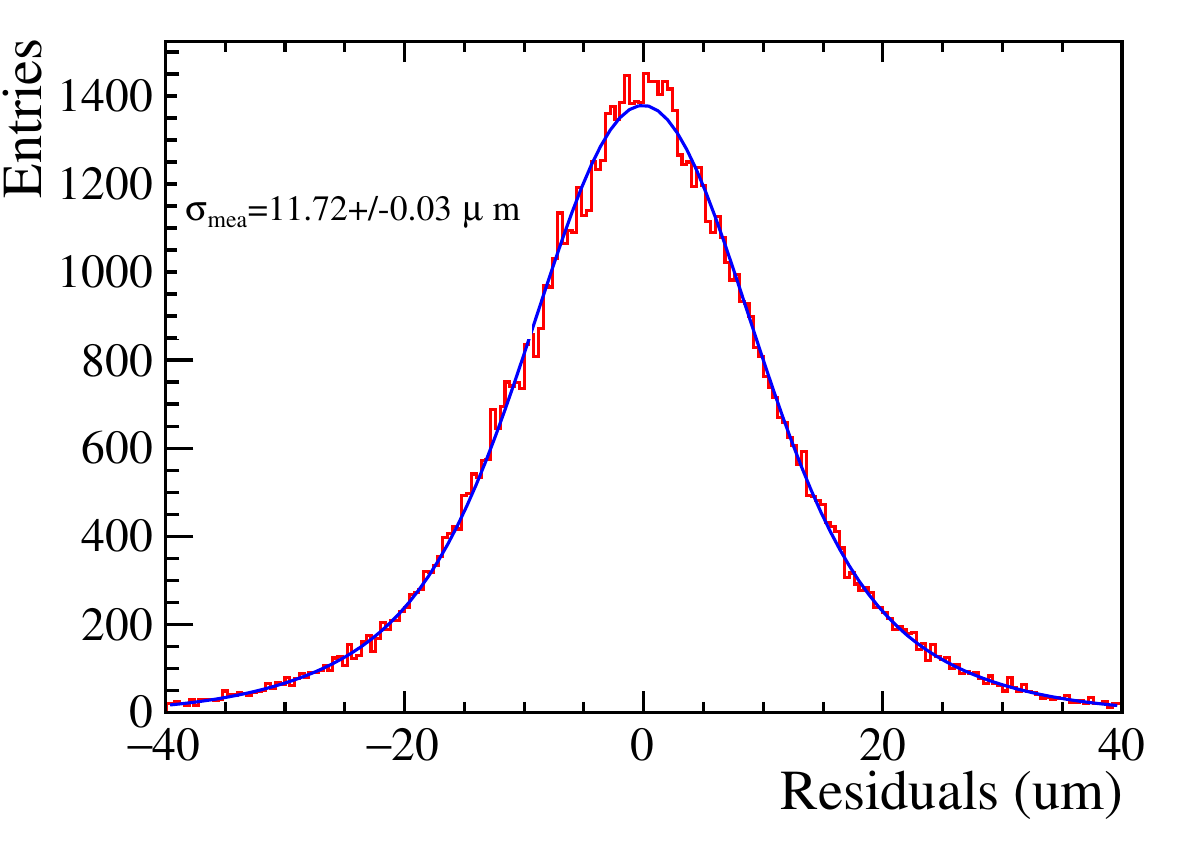}}
\hspace{0.1cm}
\subfloat[\label{fig:spatial_vs_SSD}]
{\includegraphics[width=0.48\hsize]{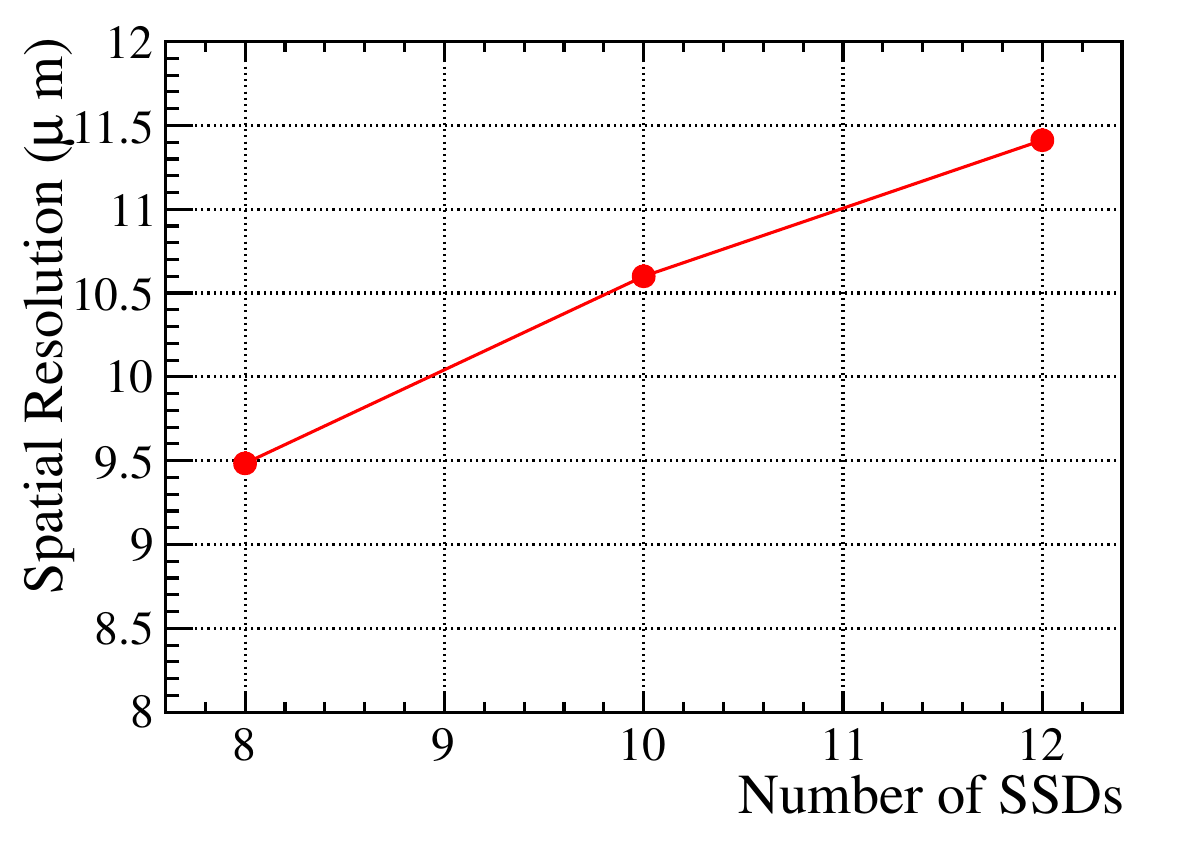}} 
\caption{(a) (b) (c) Measured residual distributions for the 8-, 10-, and 12-SSD ladders, respectively. The $\sigma_{\mathrm{mea}}$ values, obtained from double-Gaussian fits, are indicated in each panel. (d) Intrinsic spatial resolution ($\sigma_{\mathrm{DUT}}$) as a function of the number of SSDs after subtracting the telescope contribution.}
\end{figure}

For each reconstructed cluster, the hit position is determined using the developed double-$\eta$ algorithm~\cite{Miao:2025ldv}. By fitting the particle trajectory with the telescope, the predicted hit position at the DUT can be obtained. The residual is defined as the difference between the hit position predicted by the telescope and the hit position reconstructed by the DUT. The residual distributions for the 8-, 10-, and 12-SSD ladders are shown in Fig.~\ref{fig:8SSD_residual}, Fig.~\ref{fig:10SSD_residual}, and Fig.~\ref{fig:12SSD_residual}, respectively. The residual distribution is fitted with a double-Gaussian function, and the weighted width is taken as the measured spatial resolution, denoted as $\sigma_{\mathrm{mea}}$.

Based on the single-layer telescope resolution and the detector geometry, the pointing resolution of the telescope extrapolated to the DUT position, $\sigma_{\mathrm{tel}}$, can be calculated~\cite{Li2024MCS}. The intrinsic spatial resolution of the DUT, $\sigma_{\mathrm{dut}}$, is then obtained by subtracting the telescope contribution in quadrature:
\begin{equation}
\sigma_{\mathrm{dut}} = \sqrt{\sigma_{\mathrm{mea}}^{2} - \sigma_{\mathrm{tel}}^{2}} \, .
\end{equation}

Figure~\ref{fig:spatial_vs_SSD} shows the intrinsic spatial resolution for ladders with different numbers of SSDs. As the number of SSDs increases, the noise level rises while the signal amplitude remains essentially unchanged, leading to a gradual degradation of the S/N. Consequently, the intrinsic spatial resolutions for the 8-, 10-, and 12-SSD ladders are
$9.5\,\mu\mathrm{m}$, $10.6\,\mu\mathrm{m}$, and $11.4\,\mu\mathrm{m}$, respectively. 

\subsection{Comparison of \emph{Head} and \emph{Tail} SSD}

For the Layer-0 ladders, multiple SSDs are connected in series to form strips with an effective length of $\sim \! 1 \mathrm{m}$, which, to our knowledge, represents the longest single detection unit constructed to date. It is therefore worthwhile to verify whether a consistent response can be maintained along the entire strip length.

\begin{figure}[ht]
\centering
\includegraphics[width=0.8\hsize]{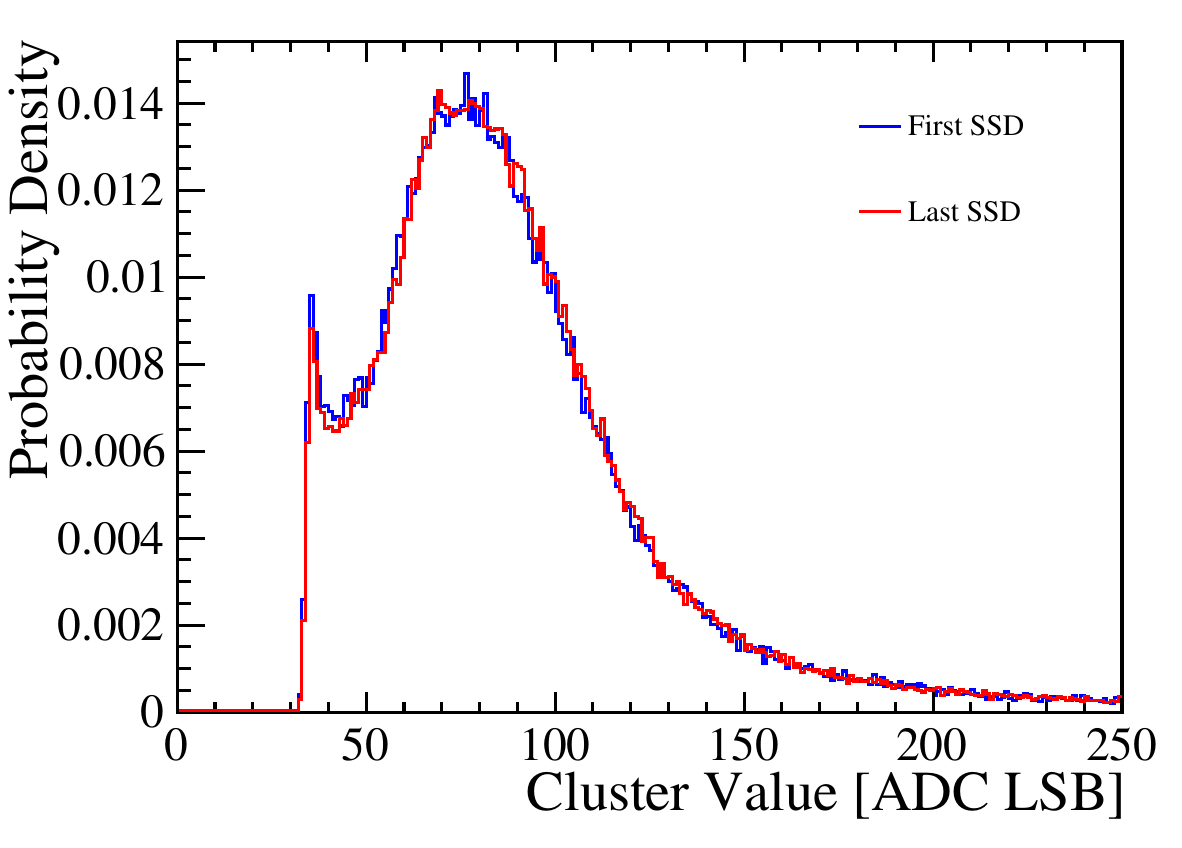}
\caption{The cluster value distributions measured with the beam hitting on the \emph{Head} and \emph{Tail} SSDs for a 12-SSD ladder. \label{fig:firt_last_mpv}}
\end{figure}

We investigate it by hitting the beam on the \emph{Head} and \emph{Tail} SSDs, and the resulting cluster value distributions are shown in Fig.~\ref{fig:firt_last_mpv}. The cluster value distributions measured at the \emph{Head} and \emph{Tail} SSDs are almost identical. In particular, no significant reduction of the signal amplitude is observed at the \emph{Tail} position compared to the \emph{Head}, indicating that there is no appreciable signal loss during the charge transport along the long strips to the front-end readout electronics.



\subsection{Performance with the incidence angle}

Since cosmic rays arrive from all directions, the detector performance is of interest not only for normal incidence but also for particles impacting at different angles. Taking into account the acceptance of AMS-02, data were collected at five incidence angles of $0^\circ$, $10^\circ$, $20^\circ$, $25^\circ$, and $30^\circ$. In all configurations, the beam was directed onto the \emph{Middle} SSD of a 12-SSD ladder.

\begin{figure}[ht]
\centering
\subfloat[\label{fig:mean_cluster_size}]{\includegraphics[width=0.48\hsize]{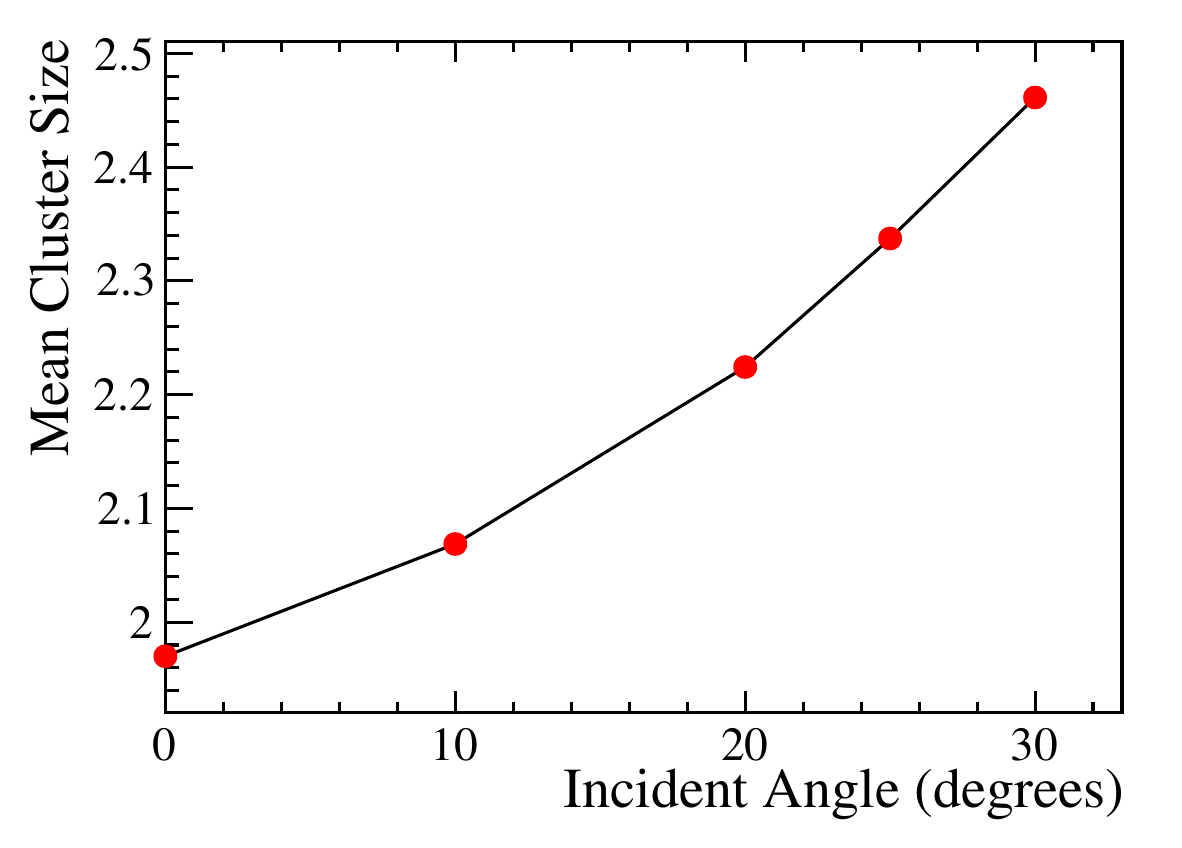}}
\hspace{0.1cm}
\subfloat[\label{fig:clsuter_size_fraction}]
{\includegraphics[width=0.48\hsize]{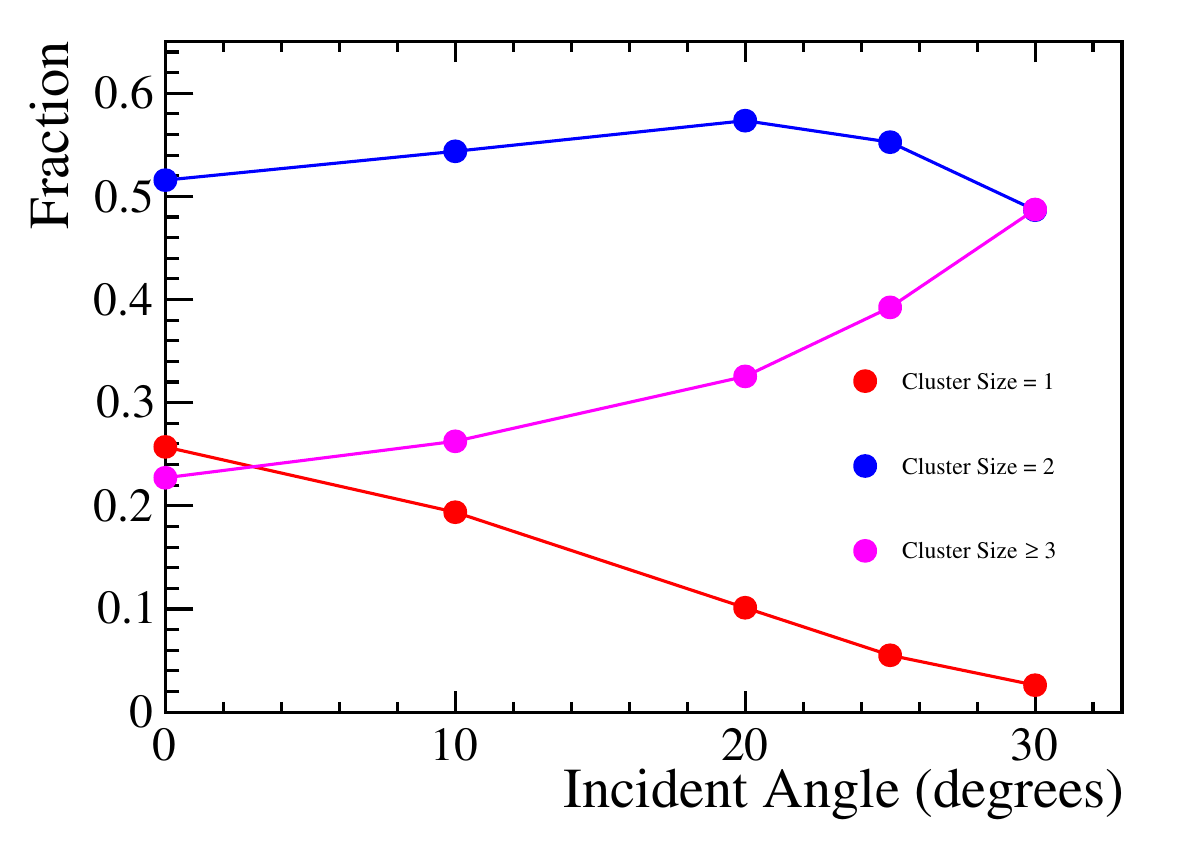}} 
\caption{(a) Mean cluster size as a function of the incidence angle. (b) Fractions of events with cluster size equal to 1, 2, and $\geq 3$ as a function of the incidence angle.}
\end{figure}

We first examine the dependence of the cluster size on the incidence angle. As the incidence angle increases, the particle traverses a longer path inside the silicon sensor, resulting in a larger total energy deposition. As electron--hole pairs are generated along the full particle trajectory, inclined tracks result in charge carriers being distributed over a larger number of readout strips. Consequently, the cluster size increases gradually with increasing incidence angle. As shown in Fig.~\ref{fig:mean_cluster_size}, the average cluster size exhibits a clear increase with the incidence angle, from about 1.97 at normal incidence to around 2.46 at 30$^\circ$.

The relative fractions of clusters with different sizes directly influence the achievable spatial resolution. For events with cluster size equal to 1, the hit position cannot be determined by interpolation between neighboring strips, and the spatial resolution is limited to the binary resolution, given by the readout pitch divided by $\sqrt{12}$. Events with cluster size equal to 2, however, enable charge-weighted interpolation between two adjacent strips and thus yield significantly better position accuracy~\cite{Wang:2000ki}. For the SSDs used in this work, the binary resolution corresponds to $109/\sqrt{12} \approx 31.5~\mu\mathrm{m}$. Three floating strips are implemented between adjacent readout strips to enhance charge sharing, thereby increasing the fraction of two-strip clusters and reducing the fraction of single-strip clusters. As shown in Fig.~\ref{fig:clsuter_size_fraction}, the fraction of clusters with size equal to 2 reaches about 50\% at normal incidence, enabling a spatial resolution of approximately $11.4~\mu\mathrm{m}$, which is much better than that achieved in the binary case.

\begin{figure}[ht]
\centering
\includegraphics[width=0.8\hsize]{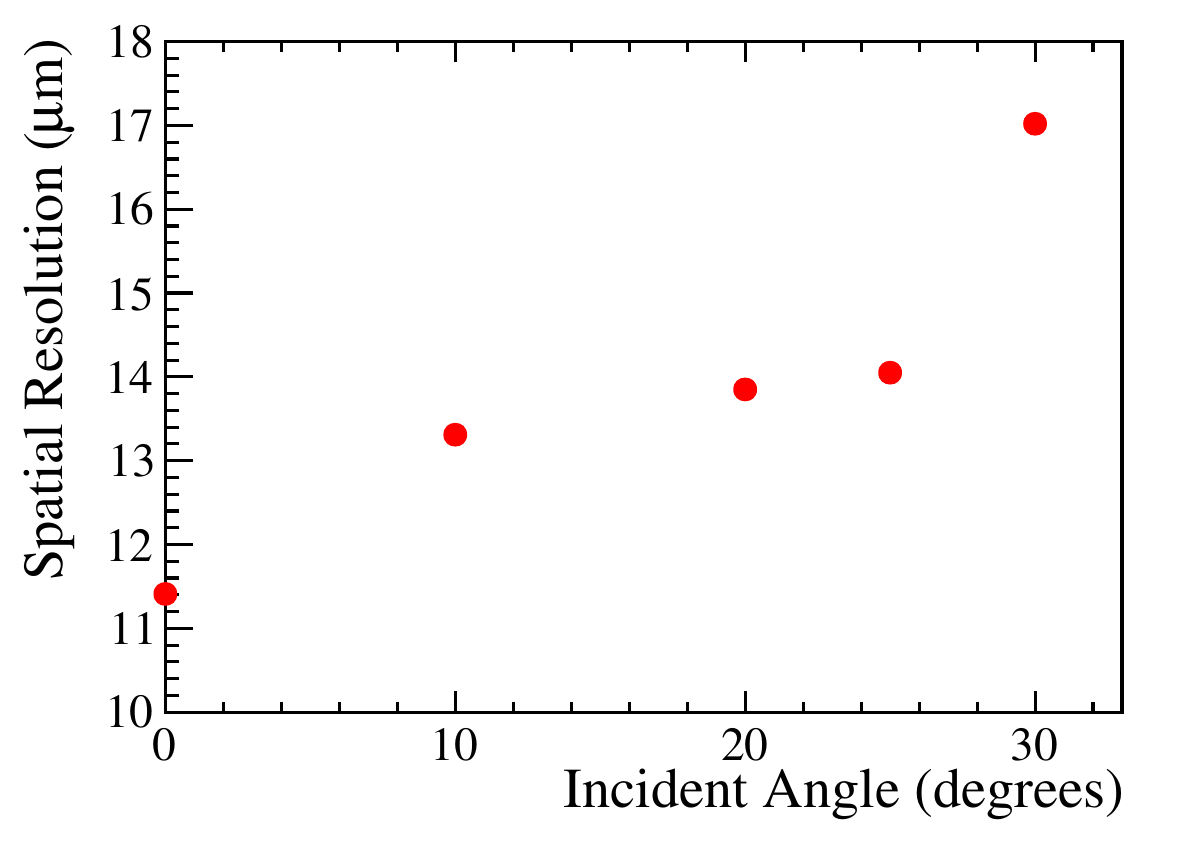}
\caption{Intrinsic spatial resolution of a 12-SSD ladder as a function of the incidence angle, measured with the beam impacting on the \emph{Middle} SSD.\label{fig:res_vs_angle}}
\end{figure}

As the incidence angle increases, the fraction of clusters with size $\geq 3$ rises, as shown in Fig.~\ref{fig:clsuter_size_fraction}, and becomes the same as that of cluster size $=$ 2 at an incidence angle of 30$^\circ$. Only two readout channels are sufficient to achieve good spatial resolution in hit position reconstruction. In contrast, clusters with size $\geq 3$ tend to spread the signal over multiple strips, thereby reducing the signal-to-noise ratio of each individual strip and ultimately limiting the position reconstruction accuracy~\cite{Turchetta:1993vu}. 

This effect is also observed in our measurements. Fig.~\ref{fig:res_vs_angle} shows the spatial resolution as a function of the incidence angle. As the incidence angle increases, the growing fraction of clusters with size $\geq 3$ leads to a gradual degradation of the spatial resolution, which reaches about $17~\mu\mathrm{m}$ at an incidence angle of 30$^\circ$. In addition, since the energy deposition of charged particles is not perfectly uniform, Landau fluctuations of the energy loss along the particle trajectory for inclined tracks further limit the achievable spatial resolution.

\section{Summary}\label{sec5}

In this work, we presented a test-beam study of the flight-model silicon microstrip detector ladders developed for the Layer-0 upgrade of the AMS-02 experiment. The study focused on the performance of ladders with different numbers of SSDs, the consistency of the detector response along the ladder, and the dependence of detector performance on the particle incidence angle.

The intrinsic noise level increases with the number of SSDs in the ladder, while the signal amplitude remains essentially unchanged. The intrinsic spatial resolutions are measured to be $9.5\,\mu\mathrm{m}$, $10.6\,\mu\mathrm{m}$, and $11.4\,\mu\mathrm{m}$ for the 8-, 10-, and 12-SSD ladders at normal incidence, respectively. No significant signal loss is observed between the \emph{Head} and \emph{Tail} SSDs of a 12-SSD ladder. The spatial resolution degrades from $11.4\,\mu\mathrm{m}$ at normal incidence to approximately $17\,\mu\mathrm{m}$ at an incidence angle of $30^\circ$.

Overall, the performance of the FM ladders satisfies the requirements of the AMS-02 Layer-0 detector upgrade. The assembled Layer-0 plane is currently undergoing several kinds of performance tests and is expected to be mounted on the International Space Station in the near future.


\section*{Acknowledgments}
This study was supported by the National Key Programme for S\&T Research and Development (Grant NO.: 2022YFA1604800), and the National Natural Science Foundation of China (Grant NO.: 12342503). We express our gratitude to our colleagues in the CERN accelerator departments for the excellent performance of the SPS.

\bibliography{reference}
\bibliographystyle{elsarticle-num}

\end{document}